# IMPACT OF LIMITED FEEDBACK ON MIMO-OFDM SYSTEMS USING JOINT BEAMFORMING

NAJOUA ACHOURA[1] AND RIDHA BOUALLEGUE[2]

[1]Department National Engineering School of Tunis, Tunisia
`najoua.achoura@gmail.com`
[2] SUP'COM, 6'Tel Laboratory,Tunisia
`ridha.bouallegue@supcom.rnu.tn`

## ABSTRACT

In multi input multi output antenna systems, beamforming is a technique for guarding against the negative effects of fading. However, this technique requires the transmitter to have perfect knowledge of the channel which is often not available a priori. A solution to overcome this problem is to design the beamforming vector using a limited number of feedback bits sent from the receiver to the transmitter. In the case of limited feedback, the beamforming vector is limited to lie in a codebook that is known to both the transmitter and receiver.When the feedback is strictly limited, important issues are how to quantize the information needed at the transmitter and how much improvement in associated performance can be obtained as a function of the amount of feedback available.In this paper channel quantization schema using simple approach to codebook design (random vector quantization)is illustrated. Performance results show that even with a few bits of feedback, performance can be close to that with perfect channel knowledge at the transmitter.

## 1. INTRODUCTION

Multiple-input multiple-output (MIMO) systems, which use multiple antennas at both transmitter and receiver, provide spatial diversity that can be used to mitigate signal-level fluctuations in fading channels [2].When the channel is unknown to the transmitter, diversity can be obtained by using space-time codes [3],[2].When channel state information (CSI) is available at the transmitter, however, diversity can be obtained using a simple approach known as transmit beamforming and receive combining. Compared with space-time codes, beamforming achieves the same diversity order as well as additional array gain, thus it can significantly improve system performance. This approach, however, requires knowledge of the transmit beamforming vector at the transmitter. In practice, CSIT must be provided to the BS by some form of feedback.
CSIT feedback schemes are a very active area of research (see for example [13] and the special issue [14] for a fairly complete list of references). In brief, we may identify three broad families: 1) open-loop schemes based on channel reciprocity and uplink training symbols, applicable to
Time-Division Duplexing (TDD); 2) closed-loop schemes based on feeding back the unquantized channel coefficients (analog feedback); 3) closed-loop schemes based on explicit quantization of the channel vectors and on feeding back quantization bits, suitably channel-encoded (digital feedback).

When the uplink and downlink channels are not reciprocal (as in a frequency division duplexing system), this necessitates that the receiver informs the transmitter about the desired transmit beamforming vector through a feedback control channel. The beamforming techniques proposed for narrowband channels can be easily extended to frequency selective channels by employing orthogonal frequency division multiplexing (OFDM). The combination of MIMO and OFDM (MIMO-OFDM), converts a broadband MIMO channel into a set of parallel narrowband MIMO channels, one for each subcarrier [10]. Transmit beamforming can be performed independently for each subcarrier of MIMO-OFDM. In non-reciprocal channels, this means requires that the MIMO-OFDM receiver calculates and sends back to the transmitter the optimal beamforming vector for every subcarrier. Practically, the feedback rate can be managed by using limited feedback techniques where the beamforming vectors are quantized using a beamforming codebook designed for narrowband MIMO channels [17]. For MIMO-OFDM's structures different approaches are possible. In this paper, joint beamforming, that consists in the extension to the transmitter side of the classical receive beamforming, is used. We focused this analyse on the impact of limited CSI on the transmitter on the performance of such system. This paper is organized as follows. In Section 2 we present the basic system model in brief. Section 3 introduce channel quantization limited feedback model, random vector quantization is presented in section 4. Finally in Section 5 some simulation results and conclusions are presented.

## 2. SYSTEM MODEL

In adaptive Beamforming, an array of antennas is exploited to reach maximum reception in a specified direction: the idea is to estimate the signal arrival from the desired direction (in the presence of noise) while signals of the same frequency from other directions are not accepted [2]. This can be achieved by varying the weights of each of the antennas used in the array. This spatial separation aims to separate the desired signal from the interfering signals. In adaptive beamforming the optimum weights are computed using complex algorithms based upon different criteria.

The communication over a frequency selective MIMO channel with $N_T$ transmits and $N_R$ receive antenna can be represented in multi-carrier fashion as:

$$y_k = H_k s_k + n_k \quad 1 \leq k \leq N \quad (1)$$

Where k denotes the carrier index, N is the number of carriers, $s_k \in \mathbb{C}^{n_T \times 1}$ is the transmitted vector, $H_k \in \mathbb{C}^{n_R \times n_T}$ is the channel matrix, $y_k \in \mathbb{C}^{n_R \times 1}$ is the received signal vector, and $n_k \in \mathbb{C}^{n_R \times 1}$ is a zero-mean circularly symmetric complex Gaussian noise vector with arbitrary covariance matrix $R_k$. Since transmit beamforming is used at each carrier, the transmitted signal is:

$$s_k = b_k x_k \quad 1 \leq k \leq N \quad (2)$$

Where $b_k$ is the transmit beamvector and $x_k$ is the transmitted symbol at the $k^{th}$ carrier [2]. The receiver also uses beamforming:

$$\hat{x}_k = a_k^H y_k \quad 1 \leq k \leq N \quad (3)$$

Where $a_k \in \mathbb{C}^{n_R \times 1}$ is the receive vector and $x_k$ is estimated symbol at the $k^{th}$ carrier. The transmitted is constrained in its average transmit power as:

$$\sum_{k=1}^{N} E\{\|s_k\|^2\} = \sum_{k=1}^{N} \|b_k\|^2 \leq P_t \quad (4)$$

where $P_T$ is the power per block-transmission. Employing limited feedback in coherent MIMO-OFDM communication systems requires cooperation between the transmitter and receiver. At the reception, the estimate of the forward link channel matrix H is used to design feedback that the transmitter uses to adapt the transmitted signal to the channel.

There are two approaches to design feedback: quantizing the channel or quantizing properties of the transmitted signal. For most closed-loop schemes, either method can be employed. It will be apparent, however, that channel quantization offers an intuitively simple approach to closed-loop MIMO, but lacks the performance of more specialized feedback methods

## 3. CHANNEL QUANTIZATION

The basic idea behind closed-loop MIMO is to adapt the transmitted signal to the channel. One approach to limited feedback is to employ channel quantization, which is illustrated in Fig. 1. This problem is reformulated as a Vector Quantization problem (VQ) by stacking the columns of the channel matrix H into a $Mr \times Mt$ dimensional complex vector $h_{vec}$. The vector $h_{vec}$ is then quantized using a VQ algorithm.

A vector quantization works by mapping a complex valued vector into one of a finite number of vector realizations. The mapping is usually designed to minimize some sort of distortion function such as the average mean squared error (MSE) between the input vector and the quantized vector.

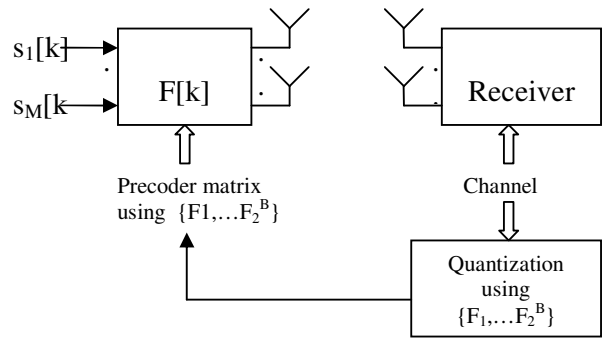

Figure 1. Limited feedback linear precoded MIMO system

Sending a quantized version of the forward link channel from receiver to transmitter gives the transmitter more flexibility to choose among different space-time signaling techniques.
Intuitively, one might expect that a random selection of matrices in the codebook F is likely to result in a large subspace distance between any pair of matrices in the codebook. This intuition is valid for a large number of antennas Mt, and is related to the fact that two vectors with i.i.d. components become orthogonal (with probability one) as the length becomes large. In the case of a random MIMO channel with i.i.d. components, the columns of the optimal precoding matrix are eigenvectors of the channel covariance matrix, which are isotropically distributed. These considerations motivated the Random vector quantization (RVQ) scheme proposed in [7], in which the elements of the codebook F are independently chosen random unitary matrices (i.e., Fk*Fk = I for each k).When used for beamforming in a MISO channel, RVQ is asymptotically optimal in the sense that it achieves the maximum rate over any codebook.

Furthermore, the achievable rate can be computed for both MISO and MIMO channels [7]. Here asymptotic means for a large system in which the number of antennas Mr and Mt each go to infinity with fixed ratio (or in the MISO case Mt goes to infinity), while also fixing B/MtM, the number of feedback bits per dimension.

### 3.1. Digital Channel Feedback Model

At the beginning, each receiver quantizes its channel to B bits and feeds back the bits perfectly to the access point. When each mobile has a single antenna (N=1), vector quantization is performed using a codebook C that consists of $2^B$ M-dimensional unit norm vectors $C = \{w_1, \cdots w_{2^B}\}$. Each receiver quantizes its channel vector to the quantization vector that forms the minimum angle to it [4] [5]. Thus, user ι quantizes its channel to $\hat{h}_i$, chosen according to:

$$\hat{h}_i = \arg \min_{w = w_1, \cdots, w_{2^B}} \sin^2(\angle(h_i, w))$$

and feeds the quantization index back to the transmitter.

In this work we use random vector quantization (RVQ), in which each of the $2^B$ quantization vectors is independently chosen from the isotropic distribution on the M-dimensional unit sphere [6]. Each receiver is assumed to use a different and independently generated codebook, and we analyze performance averaged over the distribution of random codebooks.

When N > ι, the quantization codebook consists of matrices and the distance metric can be appropriately defined. Furthermore, random quantization corresponds to choosing the quantization matrices independently from the set of all unitary matrices.

### 3.2. Linear Precoding

After receiving the quantization indices from each of the mobiles, the AP uses linear precoding to transmit data to the mobiles. When N = 1, we consider the simple strategy of zero-forcing beamforming (ZFBF). Since the transmitter does not have perfect CSI, ZFBF is performed based on the quantizations instead of the channel realizations. When ZFBF is used, the transmit vector is defined as $x = \sum_{i=1}^{M} v_i s_i$ where each σι is a scalar (chosen complex Gaussian with power P/M) intended for the ι-th receiver, and and vι ∈ ℂM is the beamforming vector for the ι-th receiver. The beamforming vectors v1,…, vM are chosen as the normalized rows of the matrix $[\hat{h}_i, \ldots, \hat{h}_M]^{-1}$, and thus they satisfy $\|v_j\| = 1$ for all $\hat{h}_i^H v_j = 0$ for all φ ≠ ι. The resulting SINR at the ι-th mobile is [8]:

$$SINR_i = \frac{\frac{P}{M}|h_i^H v_i|^2}{1 + \sum_{j \neq i} \frac{P}{M}|h_i^H v_j|^2}$$

The achievable long-term average rate is the expectation of log (1+ΣINPι) over the distribution of the fading and RVQ. When N > ι, ZFBF can be generalized to block diagonalization [9].

## 4. RANDOM VECTOR QUANTIZATION

In [12], authors analyzed, with limited channel knowledge at the transmitter, the channel capacity with perfect channel knowledge at the receiver,. Specifically, the optimal beamformer is quantized at the receiver, and the quantized version is relayed back to the transmitter. Given the quantization codebook $C = \{w_1, \cdots w_{2^B}\}$, which is also known a priori at the transmitter, and the channel H, the receiver selects the quantized beamforming vector to maximize the instantaneous rate, [11]

$$w(H) = \arg\max_{v_j \in V} \left\{ \log(1 + \rho \|Hw_j\|^2) \right\}$$

where $\rho = 1/\sigma_n^2$ is the background signal-to-noise ratio (SNR). The (uncoded) index for the rate-maximizing beamforming vector is relayed to the transmitter via an error-free feedback link. The capacity depends on the beamforming codebook V and B. With unlimited feedback (B→∞) the w(H) that maximizes the capacity is the eigenvector of H*H, which corresponds to the maximum eigenvalue.

We will assume that the codebook vectors are independent and isotropically distributed over the unit sphere. It is shown in [12], that this RVQ scheme is optimal (i.e., maximizes the achievable rate) in the large system limit in which (B,Nt,Nr)→∞ with fixed normalized feedback B = B/Nt and ¯Nr = Nr/Nt. (For the MISO channel Nr = 1). Furthermore, the corresponding capacity grows as log(ρNt), which is the same order-growth as with perfect channel knowledge at the transmitter. Although strictly speaking, RVQ is suboptimal for a finite size system, numerical results indicate that the average performance is often indistinguishable from the performance with optimized codebooks [12], [14].

## 5. SIMULATIONS RESULTS:

In this section, we evaluate the impact of working with limited feedback; witch is more practical, on system's performance, especially on the bit error rate. We consider only the case of one user (mono-user system). That means there is no need to user selection algorithms and the is no interference .Fig 2 shows the system's performance when there is two antenna in the transmitter and only one antenna on the reception (MISO system).

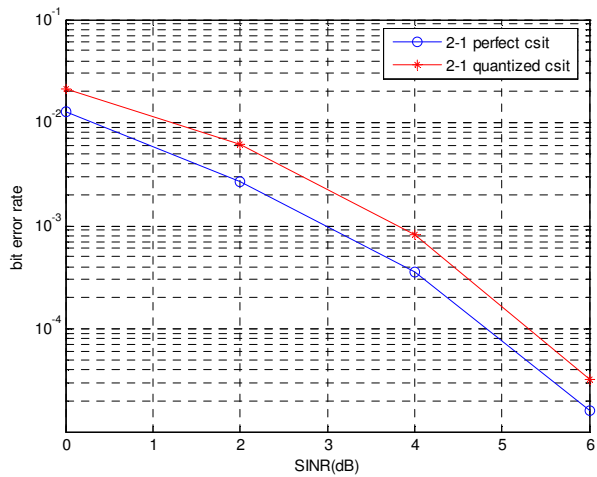

Figure 2. Nt=2, Nr=1 (with perfect channel knowledge)

Results show that for Nt = 2 and Nr=1, the Bit error rate with perfect channel knowledge at both the transmitter and receiver is larger than the rate with random vector quantified feedback. And this is perfectly expected so the blue curve is considered as the ideal case

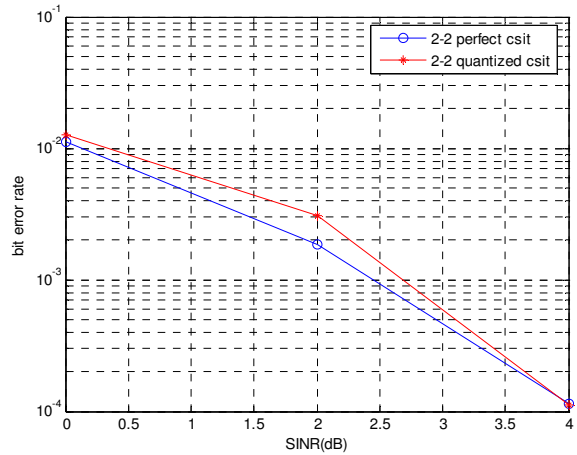

Figure 3.  Nt=2, Nr=2

In Figure 3, we consider MIMO system with two antennas both in the transmitter and the receiver, using the same parameters. Results are better here so red curve witch represents quantified feedback is more close to the ideal case.

In other hand, we consider the case where we use channel estimation in figure 4, we consider the case when all training sequences are dedicated for estimation i.e. there is no data blocs. We show estimation's result for orthogonal phase shift sequences,

In fact we can easily remark degradation of performance between the two figures, this of course can be explained by the addition of two types of error: error due to channel estimation transmitted by feedback and the second error called quantization error.

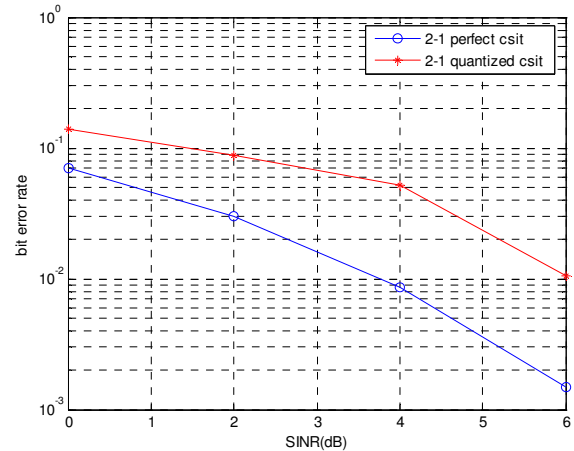

Figure 4.  Nt=2, Nr=1 (with channel estimation)

## **3. CONCLUSIONS**

This paper outlines a general framework for enabling limited feedback in closed-loop MIMO-OFDM systems. We review the application of limited feedback to MIMO communication. Numerical examples illustrate that relatively little feedback can provide substantial performance improvements.

Channel estimation error and channel evolution will definitely compromise expected performance improvements, but simulations and experimental results are required to determine how "recent" the feedback bits must be to maintain satisfactory performance.

More work is also needed in the area of limited feedback applications in MIMO-OFDM systems. In fact, a more practical technique is to feed back information on a select subset of tones and then use interpolation techniques. Other applications of limited feedback such as for multi-user MIMO channels are promising areas for investigation.